\documentclass{aa}
\usepackage{graphics}

\begin{document}

\thesaurus{06
	(08.09.2 SGR 1806-20; 
	 08.16.1; 
	 08.16.7; SGR 1806-20 
	 13.07.1; 
	 13.09.6; 
	 13.25.5 
)} 

\title{ISO observations of the environment of the soft gamma-ray\\ 
repeater SGR~1806-20}
   \titlerunning{ISO observations of the environment of SGR~1806-20}

\author{Y.~Fuchs$^{1}$, F.~Mirabel$^{1,2}$, S.~Chaty$^{1,3}$, A.~Claret$^{1}$,
 C.~J.~Cesarsky$^{1}$, D.~A.~Cesarsky$^{4}$}
	\authorrunning{Y.~Fuchs et al.}

\offprints{Y. Fuchs}
\mail{yfuchs@cea.fr}

\institute{Service d'Astrophysique, CEA Saclay, 
Bat. 709, Orme des Merisiers, 91191 Gif sur Yvette cedex, France
\and
Instituto de Astonom\'\i a y F\'\i sica del Espacio, 
cc67, suc 28. 1428 Buenos Aires, Argentina
\and
Centre d'Etude Spatiale des Rayonnements (CESR), 
9, avenue du Colonel Roche, BP 4346, 31028 Toulouse cedex 4, France
\and
Institut d'Astrophysique Spatiale, 91405 Orsay, France
}

   \date{Received 1 June 1999 / Accepted 6 September 1999}
   \maketitle

\begin{abstract}
Observations at near\footnote[5]{Based on observations collected at the European Southern Observatory, Chile (proposal ESO N\degr 59.D-0719)} and mid-infrared wavelengths (1-18\,$\mu$m) of 
SGR~1806-20 suggest that it is associated with 
a cluster of giant massive stars which are enshrouded in a dense 
cloud of dust. The centre of the best sky position 
of the gamma-ray source (Hurley et al. \cite{hurley99}) lies 
on top of the dust cloud at only 7 arcsec ($\sim$~0.5\,pc at a distance of 
14.5\,kpc) 
from the star cluster, and 12 arcsec ($\sim$~0.85\,pc) from a Luminous 
Blue Variable 
Star (LBV) which had been proposed to be associated with the SGR (Kulkarni et 
al. \cite{kulkarni95}). 
The bright cloud of interstellar gas and dust observed with ISO 
(Infrared Space Observatory) is probably 
the birth site of the cluster of massive stars, the LBV star, and 
the progenitor of the soft gamma-ray repeater pulsar. 
The presence of such a young star formation region is compatible with the 
current interpretation of soft 
gamma-ray repeaters as young neutron stars. The SGR 1806-20 compact source 
is unlikely to form a bound binary system with any of the infrared 
luminous massive stars, since no flux variations in the near-infrared 
were detected from the latter in an interval of 4 years.
The ISO observations were made over two epochs, 11 days before and 2 hours 
after a 
soft gamma-ray burst detected with the Interplanetary Network, and they show 
no enhanced mid-infrared emission associated to the high energy 
activity of the SGR. 
\end{abstract}

\section{Introduction}
 Soft Gamma-ray Repeaters (SGRs) are transient $\gamma$-ray sources that are 
distinguished from classical Gamma-Ray Bursts (GRBs) by their short duration 
($\sim$ 0.1\,s), softer $\gamma$-ray spectra and recurrent activity of their 
outbursts (Golenetskii et al. 1984). 
SGRs randomly undergo quiescent periods and intervals of 
intense activity with up to hundreds of bursts. The latter can appear as 
single short (less than 10\,ms) bursts, or complex events lasting hundreds of 
milliseconds.\\

Only four SGRs have been found so far:\\
\indent - SGR 1806-20: associated with an X-ray source coinciding with the 
radio peak of the supernova remnant (SNR) G10.0-0.3 which is a plerion: 
a non-thermal radio nebula powered by a central pulsar (Kulkarni et al. \cite{kulkarni94}). 
Recently, Hurley et al. (\cite{hurley99}) applied a statistical method 
to derive very precise location for SGRs, and found this SGR's most likely 
position significantly displaced from this radio peak. Pulsations
in the persistent X-ray flux were discovered with a period of 7.47\,s and
a spin down rate of $8.3 \cdot 10^{-11}$\,s.s$^{-1}$ 
(Kouveliotou et al. \cite{kouveliotou98a}).\\
\indent - SGR 1900+14: associated with a soft X-ray source lying close to the 
SNR G42.8+0.6. It has entered a new phase of activity in 1998 after a long 
period of quiescence. Pulsed emission with a 5.16\,s period and a secular  
spindown at an average rate of $1.1 \cdot 10^{-10}$\,s.s$^{-1}$ was detected 
(Kouveliotou et al. \cite{kouveliotou99}).\\
\indent - SGR 0525-66: a soft X-ray counterpart with an 8~s period (but this 
source appears to be time variable), lying on the northern edge of the 
SNR N49 in the Large Magellanic Cloud. No optical, infrared or radio point 
source counterpart was found (van~Paradijs et al. \cite{van paradijs}; Fender et al. \cite{fender}).\\
\indent - SGR 1627-41: near SNR G337.0-0.1 and recently discovered 
(Kouveliotou et al. \cite{kouveliotou98b}). It indicates weak evidence of 6.7\,s pulsations (Dieters et al. \cite{dieters}).\\

The association between SGR and young (less than 10\,000 years old) SNR, 
and the detection of pulsations strongly support the argument that 
SGR sources are young neutron stars (Kouveliotou et al. \cite{kouveliotou98a}). However, 
their rarity suggests unusual physical characteristics: if the secular 
spindown is due to magnetic dipole radiation, then the corresponding dipolar 
magnetic field would be in excess of $10^{14}$ Gauss (Kouveliotou et al. \cite{kouveliotou99}). Thus these 
young neutron stars are called `magnetars' 
(Thompson \& Duncan \cite{thompson}). 
They are different from normal 
pulsars; for example their X-ray and particle emissions are powered not by 
rotation but by decaying magnetic field (Kulkarni \& Thompson \cite{kulkarni98}). 
This involves both internal heating and seismic activity that shakes
the magnetosphere and accelerates particles. This gradual release of
energy is punctuated by intense outbursts that are more plausibly 
triggered by a sudden fracture of the neutron star's rigid crust caused by 
magnetic stresses (Kouveliotou et al. \cite{kouveliotou98a}).\\

A core-jet geometry has been 
reported by Vasisht et al. (\cite{vasisht}) and Frail et al. (\cite{frail}) 
in 3.6\,cm radio continuum images. 
Since SGRs are recent remnants of massive stars, they can be surrounded by
dust and gas. 
If a source of X-rays and/or relativistic jets, such as SGR~1806-20, is inside 
or near interstellar gas clouds, then a large fraction of the energy radiated 
by the X-rays and/or injected in the form of relativistic particles could 
be dissipated by heating of the interstellar material. Thus thermal emission 
from dust may be expected. 
Mid-infrared observations can be used to test this possibility, and also to 
identify possible counterparts embedded in dense dust clouds.\\

\section{Previous observations of SGR~1806-20}

\begin{figure*}[!hp]
\centering
\resizebox{13.7cm}{!}{\includegraphics*{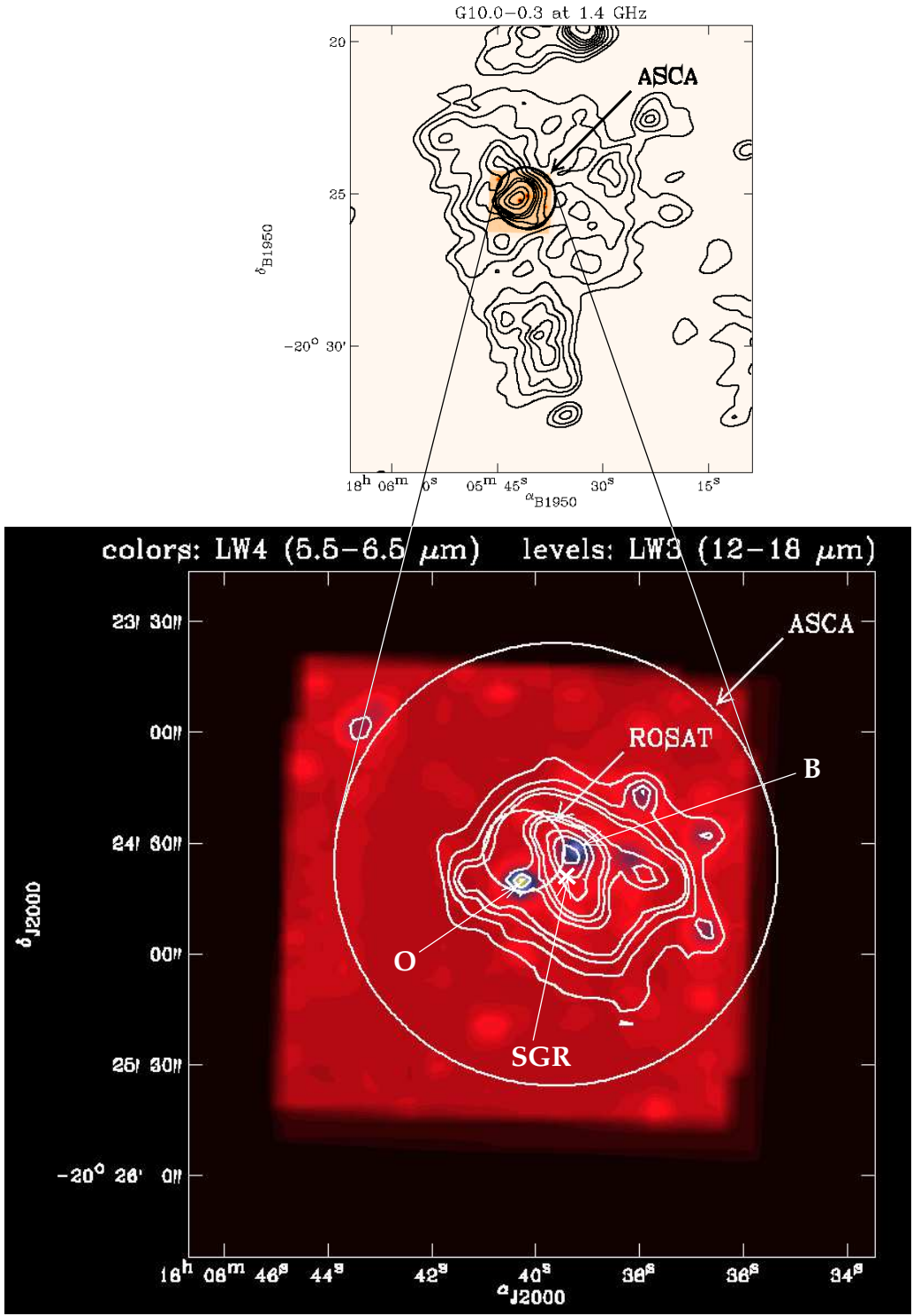}}
\caption{
{\bf Top:} Radio image of the supernova remnant G10.0-0.3 at 
1.4\,GHz (from Kulkarni et al. \cite{kulkarni94}) with the ISOCAM 
LW4 (5.5-6.5\,$\mu$m) color image superimposed on it.
The circle marks the location of the steady ASCA X-ray source 
(Murakami et al. \cite{murakami}) considered as the counterpart of 
SGR~1806-20. 
{\bf Bottom:} The best fit position of SGR~1806-20 
(Hurley et al. \cite{hurley99}) marked as a white cross, the 1\arcmin \  ASCA
circle and the 11\arcsec \  localization circle of the X-ray source by ROSAT, 
are shown superimposed on our observation of this region with ISOCAM. 
The levels of the LW3 (12-18\,$\mu$m) image are superimposed on the LW4 
(5.5-6.5\,$\mu$m) color image, both have $3\times 3$ arcsec$^2$ per pixel 
field of view. 
Contour levels are: 15, 17, 19, 20, 30, 35, 41, 61, 81, 127 mJy, and the 
images were smoothed using a bilinear interpolation. 
A Luminous Blue Variable (LBV) star, named ``O'', is on the south part of 
the ROSAT circle, and to the west, there is a cluster of giant 
massive stars, named ``B'' (see Fig.~\ref{fig2}).
Note that the coordinates drawn on this image are not precise.}\label{fig1} 
\end{figure*}

SGR 1806-20 is the most prolific soft gamma-ray repeater with more than a 
hundred bursts (Hurley et al. 1994)
since its discovery (Atteia et al. 1987; Laros et al. 1987). 
Within its localization box, Kulkarni 
and Frail (\cite{kulkarni93}) pointed out an amorphous radio nebula G10.0-0.3 
classified as a supernova remnant. Conducting a multi-band radio 
observation of this SNR using the Very Large Array (VLA), Kulkarni 
et al. (\cite{kulkarni94}) 
found a compact nebula superimposed on an extended plateau of emission. 
Cooke (\cite{cooke}), with the ROSAT X-ray telescope, discovered an 
X-ray source coincident with this compact nebula within an 11\arcsec \  error 
box. This was confirmed by Murakami  et al. (\cite{murakami}), 
who were fortunate to 
detect an X-ray burst, simultaneously detected by BATSE as a $\gamma$-ray 
burst, and a point-like source (designated AX\,1805.7-2025) 
in the same observation with the X-ray satellite ASCA. The burst 
was coincident with the steady source and both were coincident with the 
centroid of the compact radio nebula. 
This SNR (see top of Fig.~\ref{fig1}) has strong emission from the central 
region, and a hierarchical structure culminating in a central peak, suggesting 
a compact object located at this radio core.
This morphology is typical of a plerion, where the radio emission is 
synchrotron radiation powered by the relativistic wind of a central pulsar. 
In addition, Murakami et al. (\cite{murakami}) argued that the radio and the 
X-ray spectra are consistent with  a synchrotron source such as the Crab 
nebula, and the similarity of the radio/X-ray flux ratio is consistent with 
a source powered by a pulsar from radio to X-ray wavelengths. The pulsar 
model for SGR 1806-20 was confirmed later with the discovery of 7.47\,s 
pulsations in its persistent X-ray flux (Kouveliotou et al. \cite{kouveliotou98a}). 
Still, the measured period increase implies a magnetic field 
stronger than $10^{14}$ Gauss, and the rotational energy of the neutron 
star with its present period is too small to power the X-ray and particle 
emission (Kouveliotou et al. \cite{kouveliotou98a}), so the magnetar model becomes more plausible. \\

\begin{figure*}[!ht]
\centering
\resizebox{13cm}{!}{\includegraphics{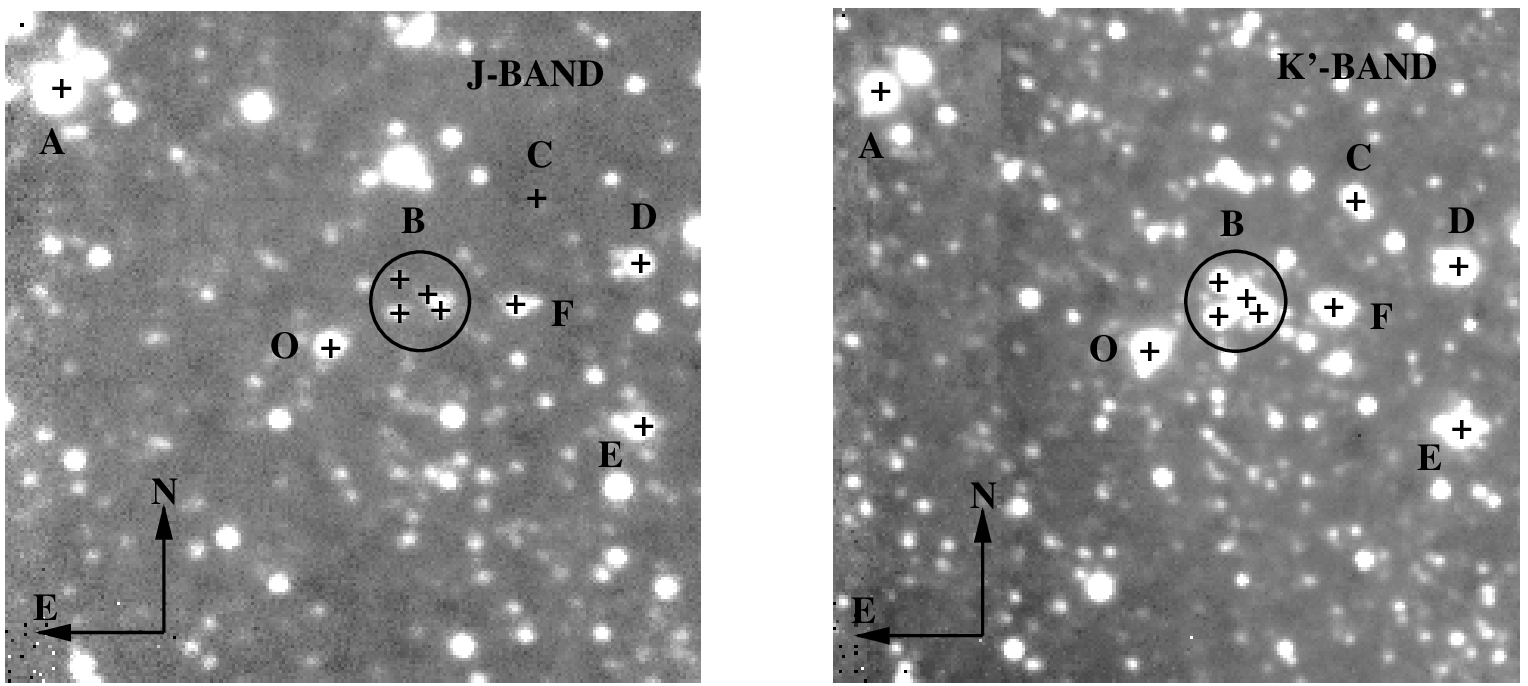}}
\caption{J-band ($1.25 \pm 0.30 \,\mu$m) and 
K$'$-band ($2.15 \pm 0.32 \,\mu$m) 
images of the field surrounding SGR~1806-20 obtained with IRAC2b on the 
ESO/MPI 2.2 m telescope with a $0.5 \times 0.5$ arcsec$^2$ per pixel field of 
view. The global field shown is roughly the same as in the ISOCAM images: 
$\sim 2' \times 2'$. The objects clearly 
seen with ISOCAM are marked. The LBV is marked ``O''. Except for ``A'', the 
extinction is greater in the J-band than in the K$'$-band. The extinction of 
``B'' in the J-band image suggests that it is a cluster of hot giant massive 
stars behind or still partly embedded in their ``placental'' dusty cloud, 
visible in the LW9 (14-16\,$\mu$m) and LW3 (12-18\,$\mu$m) images in 
Fig.~\ref{fig4}.
}\label{fig2}
\end{figure*}

Thanks to the precise location of SGR 1806-20 from the radio data, it was 
possible to search for optical or infrared counterparts. Kulkarni et al. 
(\cite{kulkarni95}) detected a highly reddened luminous supergiant star coinciding 
with the radio peak of G10.0-0.3, whose reddening is consistent with 
the high extinction inferred for AX\,1805.7-2025 (A$_{\rm V} \sim 30$\,mag). 
Spectroscopic observations by van~Kerkwijk et al. (\cite{van kerkwijk}) 
classified this star as a Luminous Blue Variable (LBV) candidate of spectral 
type O9-B2. LBVs are among the most luminous and massive stars. 
They can show no detectable variability ($\Delta$M~$<$~0.1\,mag), 
or either they can be very variable ($\Delta$M~$>$~3\,mag) on different
timescales, and they eject large amount of
matter (their mass loss rate can be as high as $10^{-5}$ to $10^{-4}$ 
M$_{\odot}$.yr$^{-1}$). Only a dozen LBVs are known in the local group of 
galaxies so 
they presumably represent a very short ($< 10^5$\,yr) stage of some rare, 
massive stars' life. This peculiar LBV is one of the brightest in our Galaxy 
with a bolometric luminosity greater than $10^6$~L$_{\odot}$. Although it
is still not clear what is the physical connection between the SGR source and 
this LBV, the chance coincidence probability of this kind of object with the 
radio position is exceedingly small ($\sim 2 \cdot 10 ^{-5}$ from Kulkarni 
et al. \cite{kulkarni95}). According to Corbel et al. (\cite{corbel})
the SNR G10.0-0.3 is very likely associated with one of the brightest
H\,{\textsc II} regions in the Galaxy: W31, on the edge of a giant molecular 
cloud, at a distance of $14.5 \pm 1.4$\,kpc with a visual extinction of 
$35 \pm 5$\,mag. The LBV's characteristics are consistent with this region, 
thus leading to nearly $5 \cdot 10^6$\,L$_{\odot}$ for its luminosity. During 
the last period of activity of SGR~1806-20 in 1996 November, Castro-Tirado 
et al. (\cite{castro-tirado}) performed follow-up observations of the LBV. 
They suggested that no strong additional IR emission appears during the 
active period, nor any variation larger than 0.1\,mag in the K-band in 1\,s 
timescale can be attributed to an X/$\gamma$-ray burst. So this luminous blue 
variable does not seem to exhibit any variability greater than 0.1\,mag 
even during the bursts, which tends to 
disprove the accreting binary model for a physical link between the LBV 
and the pulsar.\\

\section{Infrared observations}

Observations of the mid-infrared environment of SGR~1806-20 were carried out 
on 1997 April 3 (with the LW2\,(5.0-8.5\,$\mu$m) filter) and April 14 
(with the other filters) by the ISOCAM instrument 
(Cesarsky et al. \cite{cesarsky}) 
aboard the Infrared Space Observatory (ISO) satellite. 
By chance, a soft gamma-ray burst was detected by the Interplanetary Network 
on 1997 April 14 (Hurley et al. \cite{hurley99}). 
So the LW2 image was taken 11 days before, and all the other ISOCAM 
observations were made only between 1.6 and 3.5 hours after this burst. 
The images were taken in the 5-18\,$\mu$m range with several wide-band 
filters (LW), with a 1.5 arcsec pixel field of view for 
LW2\,(5.0-8.5\,$\mu$m) and a 3 arcsec 
pixel field of view for the other filters. The data were reduced using the 
``CIA'' package to correct the dark current, the glitches, the flat field and 
the detector transient behaviour (see Ott et al. 1997).\\

We also monitored the field of SGR~1806-20 in the near-infrared, 
at the European Southern Observatory (ESO), with the ESO/MPI 2.2\,m telescope 
with the IRAC2b camera. The most complete observations were made 
on 19 July 1997, in the J\,($1.25 \pm 0.30 \,\mu$m), 
H\,($1.65 \pm 0.30 \,\mu$m) and K$'$\,($2.15\pm 0.32 \,\mu$m) bands. 
The IRAC2b 
camera was mounted at the F/35 infrared adapter of the telescope. 
This camera is a Rockwell 256$\times$256\,pixels Hg:Cd:Te\,NICMOS\,3 large 
format infrared array detector. It was used with the lens C, providing 
an image scale of $0.49 \mbox{ arcsec/pixel}$ and a field of $136 \times 
136 \mbox{ arcsec}^{2}$. The typical seeing for these observations was 
$1.2 \mbox{ arcsec}$.
Each image taken at la Silla is the median of 9 images exposed 
for 1 minute, four of these images offset by 30\arcsec \ to the North,
East, South and West, to allow subtraction of a blank sky.
The images were further treated by removal of the bias, the dark current,
and the flat field, and we carried out absolute and relative photometry
to look for small variations of the luminosity of SGR 1806-20. This work
was performed with the IRAF procedures, using the DAOPHOT package for
photometry in crowded fields.\\

The ISO absolute astrometry is not very precise, so we used the relative 
positions of the sources visible on our ISOCAM images to identify them 
with the ones on the J, H and K$'$ images. For the ISOCAM images displayed 
in this paper, we corrected their astrometry using the LBV's position 
given by Kulkarni et al. (\cite{kulkarni95}) as a reference for the center 
of the corresponding ISO source. However, the accuracy of the displayed 
coordinates is still about 2\arcsec.\\

\section{Results and discussion}

\begin{figure}[!htb]
\centering
\resizebox{7.7cm}{!}{\includegraphics{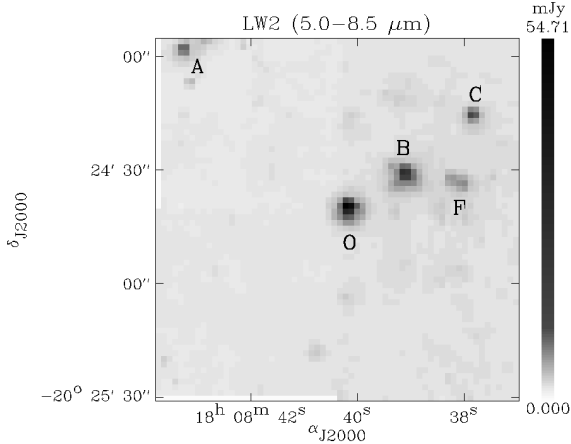}}
\caption{ISOCAM LW2\,(5.0-8.5\,$\mu$m) image of SGR~1806-20 taken with the 
$1.5 \times 1.5$~arcsec$^2$ per pixel field of view lens. 
The global field of view is $\sim 1.5' \times 1.5'$, 
smaller than the field of the other ISOCAM images ($\sim 2' \times 2'$), so 
only ``O'',``B'',``C'',``F'' and ``A'' are visible. Even with this finer 
resolution, the stars of the cluster ``B'' are not resolved.}\label{fig3}
\end{figure}

We can summarize the aforementioned information and our observations of 
SGR~1806-20 in Fig.~\ref{fig1}. The top represents the radio image of the 
SNR G10.0-0.3 at 1.4\,GHz presented by Kulkarni et al. (\cite{kulkarni94}) 
with the ASCA 1\arcmin \ radius error circle centered on the steady X-ray 
source. 
The bottom shows the same circle surrounding the ISOCAM view of this region 
with the 3 arcsec pixel field of view, 
and we overlaid the 11\arcsec \ radius localization circle of the X-ray 
source by ROSAT.
Fig.~\ref{fig2} shows J-band and K$'$-band images roughly corresponding to 
the ISOCAM field of view.  
The objects clearly seen on the ISOCAM images are marked; ``O'' is the LBV 
supposed to be the IR counterpart of SGR~1806-20, in Fig.~\ref{fig1} it is 
the bright color point on the south part of the ROSAT circle. ``B'' is a 
cluster of stars in Fig.~\ref{fig2}, but it is seen as an unresolved source 
with the lower resolution (3~arcsec per pixel) of the ISOCAM 
LW4\,(5.5-6.5\,$\mu$m) color image. Even with the 
1.5 arcsec pixel field of view, the stars of this cluster are not resolved 
between 5.0 and 8.5~$\mu$m, the range of the LW2 image (see Fig.~\ref{fig3}).\\

\begin{figure*}[!hp]
\centering
\resizebox{17.85cm}{!}{\includegraphics{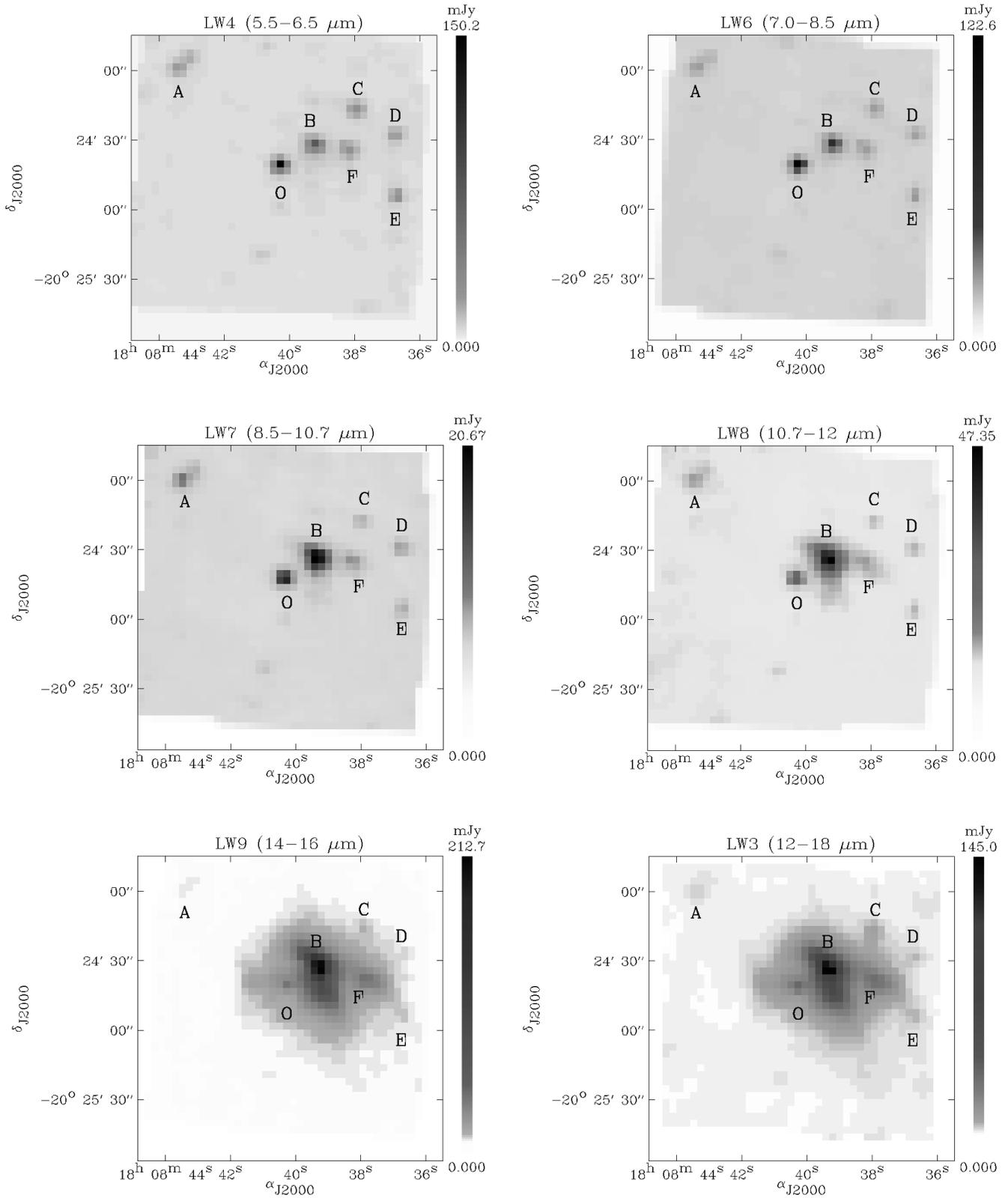}}
\caption{ 
Continuum images of the environment of SGR~1806-20 as seen by ISOCAM wide 
band filters with $3\times 3$~arcsec$^2$ per pixel field of view. 
The global field of view is $\sim 2' \times 2'$. 
At 6\,$\mu$m the LBV dominates and all the sources are point-like, 
except ``B'' (see Fig.~\ref{fig2}): the individual stars are not resolved but 
the cluster seems to
be dominated by the flux of the central star. With the increasing wavelength 
the presumed dust cloud appears and its peak seems located on the south-east 
star of the cluster (see Fig.~\ref{fig1}). Stars ``O'' (the LBV), ``C'', 
``D'', ``E'', and ``F'' seem to be in or behind this cloud.}\label{fig4}
\end{figure*}

Fig.~\ref{fig4} shows how the infrared field of view surrounding SGR~1806-20 
changes depending on the wavelength. 
At 6\,$\mu$m the LBV is the brightest star and nearly all the sources look 
point-like on the LW4 image (see Fig.~\ref{fig1} and Fig.~\ref{fig4}).
``B'' becomes brighter compared to ``O'' as the wavelength increases, and 
a cloud is clearly visible in the LW3\,(12-18\,$\mu$m) image where it strongly 
dominates the observed flux. From the superimposed LW3 level image with the 
LW4 color image in Fig.~\ref{fig1}, the cloud appears to be 
centered on the south-east star of the ``B'' cluster. From this spectral 
evolution and  the extinction of ``B'' and the other stars (except ``A'') 
in the J-band image compared to the K$'$-band one, we suggest that this is 
a dust cloud, birth place of the cluster ``B'' which is probably composed 
of hot giant massive stars. These stars are behind or still 
partially embedded in this cloud and heat it up, 
so it appears bright around 15\,$\mu$m. ``O'',``C'',``D'',``E'' and ``F'' 
seem to be in or behind this cloud, but ``A'' seems to be a foreground star.\\

\begin{figure*}[!hp]
\centering
\rotatebox{+90}{\resizebox{19.9cm}{!}{\includegraphics{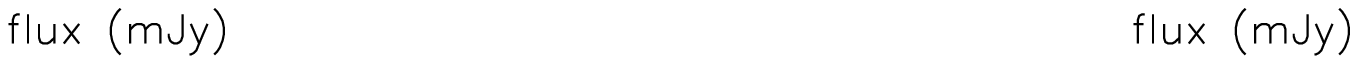}\includegraphics{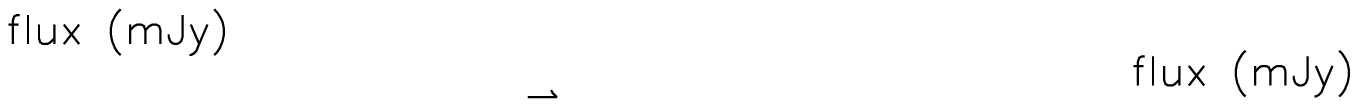}}}
\caption{Observed flux densities of ``O'', ``B'', ``C'' and ``F'' in the 
J, H and K$'$~bands, and in the large ISOCAM filter bands (see Fig.~\ref{fig4} 
for the corresponding wavelengths). We overplotted in 
each case an approximate fit for the data (solid lines) which is the sum of a 
``hot'' black body (dashed lines) and a ``cold'' one (dotted lines), both 
attenuated by the interstellar extinction law of Rieke and Lebofsky 
(1985) with  A$_{\rm V} = 30$\,mag, except for ``A''  whose 
spectrum corresponds more likely to A$_{\rm V} = 14$\,mag.
So ``A'' is a foreground star. 
For each fit, the presumed temperature and radius (in solar radius unit 
R$_{\odot}$) of the emitting region are noted down. 
``B'' is a cluster of resolved 
stars in the J, H and K$'$ images, but is unresolved in the ISOCAM images. 
So for its photometry in the near infrared, we integrated the flux density 
over a circle surrounding the four stars of this cluster.
``D'' and ``E'' are at the edge of the dust cloud showed in 
the LW3 image (see Fig.~\ref{fig4}), it may explain that they are less 
attenuated in the J, H and K$'$ bands than the others stars and thus that 
their flux densities are below the fit in these bands.}\label{fig5} 
\end{figure*} 

This assumption is supported by the spectral energy distribution of these 
objects. In Fig.~\ref{fig5} we show the observed flux 
densities in the J, H and K$'$~bands, and in the large ISOCAM filter bands 
(see Fig.~\ref{fig4} for the corresponding wavelenghts). We overplotted in 
each case an approximate fit for the data (solid lines) which is the sum of a 
``hot'' black body (dashed lines) and a ``cold'' one (dotted lines), both 
attenuated by the interstellar extinction law of Rieke and Lebofsky 
(\cite{rieke}). As we have already noticed, ``B'' is a cluster of resolved 
stars in the J, H and K$'$ images, but is unresolved in the ISOCAM images. 
So for its photometry in the near infrared, we integrated the flux density 
over a circle surrounding the four stars of this cluster. 
For $\lambda < 10 \,\mu$m the 
spectra are consistent with those of hot supergiant stars through the 
interstellar absorption A$_{\rm V} \sim 30$\,mag, except for ``A'' whose 
spectrum corresponds more likely to A$_{\rm V} \sim 14$\,mag, 
indicating that ``A'' is a foreground star. 
For $\lambda > 10 \,\mu$m, the flux of the dust cloud, presumed to be heated 
by these hot stars, is dominant. 
For ``O'',``C'' and ``F'', closer to the ``B'' cluster than ``D'' and ``E'', 
the dust cloud is fitted by a hotter black body (150~K and 160~K) than for 
``D'' and ``E'' (120~K and 130~K). 
The rough black body model fits our data well 
for ``O'',``B'',``C'' and ``F'' (see Fig.~\ref{fig5}). The absorption 
A$_{\rm V} \sim 30$\,mag appears to be the same for these stars, so
they are probably embedded in the cloud.\\

``O'' is very luminous, as expected from an LBV candidate. But the stars 
in the ``B'' cluster are also luminous. A simple estimate of their 
individual mean luminosities in our observational wavelenght range shows 
that each is comparable to the LBV's estimated luminosity in the same range.
We can compare the LW2\,(5.0-8.5\,$\mu$m) image taken 11 days before the 
1997~April~14 burst reported in Hurley et al. (\cite{hurley99}), and both the 
LW4\,(5.5-6.5\,$\mu$m) and LW6\,(7.0-8.5\,$\mu$m) images, taken only 2 hours 
after this burst. No new source appears on the latter images, and for each 
observed source, the fluxes in the three bands are consistent with each other, 
within the 20\% 
typical error due to ISOCAM photometry. 
Thus, there is no evidence of additional heating by the high energy 
activity of SGR~1806-20, before or right after this burst.\\

\begin{table*}[!htp]
	\caption[]{Chronology of the soft gamma-ray bursts and the infrared observations of SGR~1806-20 for the late 1993-1999 epoch.}
	\begin{minipage}{\hsize}
	\begin{tabular}{llll}
	
\hline
X/$\gamma$-ray & Infrared & Instrument and/or wavelenghts & References\\
\hline
1993 Sept 29 & & BATSE & Kouveliotou et al. 1993\\
 & 1993 Oct 4-5 & Palomar 1.5\,m; I\,(0.90\,$\mu$m) & Kulkarni et al. 1995\\
 & 1993 Oct 8 & Hale 5\,m; J\,(1.25\,$\mu$m), H\,(1.65\,$\mu$m), K\,(2.2\,$\mu$m), L$'$\,(3.7\,$\mu$m) & Kulkarni et al. 1995\\
1993 Oct 9 & & BATSE \& ASCA & Tanaka 1993\\
1993 Oct 11 & & ASCA & Tanaka 1993\\
 & & &\\
 & 1994 Jun 1 & Hale 5\,m; optical & Kulkarni et al. 1995\\
 & 1994 Jul 7 & ESO/MPI 2.2\,m; K$'$\,(2.15\,$\mu$m) & Chaty 1998\\
 & 1994 Jul 8 & ESO/MPI 2.2\,m; J\,(1.25\,$\mu$m), K$'$\,(2.15\,$\mu$m) & Chaty 1998\\
 & 1994 Aug 19 & UKIRT\footnote{United Kingdom Infrared Telescope, Mauna Kea, Hawaii}; K spectrum:\,2.02-2.22\,$\mu$m & van\,Kerkwijk\,et\,al.\,1995\\
 & 1994 Oct 18 & Hale 5\,m; J spectrum:\,1.19-1.33\,$\mu$m, H spectrum:\,1.54-1.78\,$\mu$m & van\,Kerkwijk\,et\,al.\,1995\\
 & & &\\
 & 1995 Sept 24 & JCMT\footnote{James Clerk Maxwell Telescope}; 800\,$\mu$m & Smith et al. 1997\\
 & 1995 Sept 25 & JCMT$^{\it b}$; 450\,$\mu$m, 800\,$\mu$m & Smith et al. 1997\\
1995 Sept 30 & & BATSE & Kouveliotou et al. 1995\\
 & & &\\
 & 1996 Jul 3 & GSCAO\footnote{German-Spanish Calar Alto Observatory} 1.2\,m; H\,(1.65\,$\mu$m), K\,(2.2\,$\mu$m) & Castro-Tirado\,et\,al.\,1998\\
1996 Oct 30-31 & & BATSE & Kouveliotou\,et\,al.\,1996a\\
1996 Nov 5-6 & 1996 Nov 6 & RXTE / TCS\footnote{Carlos S\`anchez Telescope, Observatorio del Teide (Canary Islands)} 1.5\,m; K\,(2.2\,$\mu$m) & Kouv96b\footnote{Kouveliotou et al. 1996b}/C-T98\footnote{Castro-Tirado et al. 1998}\\
 & 1996 Nov 10 & TCS$^{\it d}$ 1.5\,m; J\,(1.25\,$\mu$m), H\,(1.65\,$\mu$m), K\,(2.2\,$\mu$m) & Castro-Tirado\,et\,al.\,1998\\
1996 Nov 19 & & Ulysses \& BATSE & Hurley et al. 1996\\
1996 Nov 23 & & BATSE & Hurley et al. 1996\\
1996 Dec 30 & & IPN\footnote{Interplanetary Network, constiting of BATSE, Ulysses, and KONUS-WIND} & Hurley et al. 1999\\
1997 Jan 24 & & IPN & Hurley et al. 1999\\
 & 1997 Apr 3 & ISOCAM; 6.75\,$\mu$m & this paper\\
1997 Apr 14 & 1997 Apr 14 & IPN/ISOCAM; 6\,$\mu$m, 7.75\,$\mu$m, 9.6\,$\mu$m, 11.35\,$\mu$m, 15\,$\mu$m & Hurley\,et\,al.\,1999/this\,paper\\
 & 1997 July 19 & ESO/MPI 2.2\,m; J\,(1.25\,$\mu$m), H\,(1.65\,$\mu$m), K$'$\,(2.15\,$\mu$m) & this paper\\
1997 Aug 27 & & IPN & Hurley et al. 1999\\
1997 Sept 2 & & IPN & Hurley et al. 1999\\
 & & &\\
1998 Aug 5 & & IPN & Hurley et al. 1999\\
1999 Feb 5 & & IPN & Hurley et al. 1999\\
\hline
	\end{tabular}
	\end{minipage}
\end{table*}
 
Concerning the LBV ``O'' flux, our observed 
near-infrared magnitudes are: $J$~$=$~$13.72 \pm 0.15$, $H = 10.49 \pm 0.13$ 
and $K' = 8.76 \pm 0.12$.
If we compare these to our previous observations in July 1994
(Chaty \cite{chaty}), the flux variation in the J and K$'$~bands is less than 0.1\,mag. 
These values and the mid-infrared ones are consistent with 
the previous published infrared measurements (see Kulkarni et al. 
\cite{kulkarni95}; Castro-Tirado et al. \cite{castro-tirado}, and Smith et al. \cite{smith}). 
So the J, H and K$'$ magnitudes show no significant variation greater than 
0.1\,mag over an interval of four years.
Table~1 summarizes the observation dates of the SGR~1806-20 soft gamma-ray 
bursts and its infrared counterpart during this time interval. 
We have also checked the near-infrared flux of stars ``B'' and ``F'' 
in 1994 and in 1997. Again the variation is less than 0.1\,mag, the 
typical error in the flux. So there appears to be no variable 
source in the vicinity of the SGR~1806-20 radio core.\\

Now, let us suppose that the LBV is the physical counterpart of SGR~1806-20 
and that they were born in the dust cloud at $\sim  15''$ from their present 
location. The plerions' characteristics 
imply that they fade rapidly (Kulkarni et al. \cite{kulkarni94}) 
so the maximum pulsar age 
is $t_p \sim 10^4$\,yr. Then, at a distance of 14.5\,kpc, 
the pulsar would have a minimum transverse velocity 
$v_p \sim 100$\,km.s$^{-1}$, and it would be a runaway neutron star. If the 
LBV is the donor, it is not fully understood how such a massive binary would 
have acquired this peculiar velocity without being disrupted.\\

Moreover, a recent paper by Hurley et al. (\cite{hurley99}) 
presents a new estimate of SGR~1806-20's location. 
Its most likely position would have a small displacement from the 
radio core of the SNR G10.0-0.3 coinciding with the LBV. This observation is 
consistent with our data showing no evidence of any binary activity.
Hurley et al. (\cite{hurley99})
propose that the neutron star's progenitor and the LBV initially 
formed a binary system, which became unbound following the supernova 
explosion.  
It is possible that the radio core is due to the huge mass loss of the 
LBV, while the rest of the plerion is the remnant of the progenitor's 
supernova explosion. As shown in Fig.~\ref{fig1}, the new neutron star's 
location is closer to the cluster of massive stars called ``B'' in this 
paper ($\sim$ 7\arcsec \ corresponding to $\sim$ 0.5\,pc at a distance 
of 14.5\,kpc) than to the LBV ($\sim$ 12\arcsec \ corresponding to 0.85\,pc). 
``B'' lies at the edge of the 3$\sigma$ equivalent confidence contour of 
SGR~1806-20's position, presented in Hurley et al.'s (\cite{hurley99}) 
paper, whereas the LBV is well outside this contour. 
This implies 
that the birth place of the SGR's progenitor has an equal probability of 
being in the cluster, as near the actual LBV's location. 
The argument for an alignment between SGR~1806-20 and a very luminous star 
such as an LBV, to that of other stars in the cluster is equally justified, 
since the mean luminosities of the stars are comparable with that of the LBV.\\

\section{Conclusion}

SGR~1806-20 is an X-ray pulsar, lying close to the radio core of the 
plerion G10.0-0.3 (Kulkarni et al. \cite{kulkarni94}). From the latest 
models, this neutron star is surely a magnetar (Thompson \& Duncan 
\cite{thompson}) with its unusual physical characteristics leading to the 
gamma-ray outbursts.
The previous search for a counterpart inside the ASCA and ROSAT error boxes 
led to the discovery of a luminous blue variable (LBV) star, a quite unusual 
star belonging to the brightest stars in the Galaxy, coinciding with the radio 
peak of the supernova remnant.\\

We have observed with ISOCAM the luminous blue variable star 
that was previously associated with SGR~1806-20, and a dust enshrouded cluster
of equally luminous massive stars, which heat a dust cloud that appears very
bright at 12-18\,$\mu$m.
This infrared luminous cloud was probably the formation site of 
the cluster of hot stars, the LBV, and the progenitor of SGR~1806-20.
For the region where these objects lie, there is excess emission 
at 12-18\,$\mu$m, but there is no evidence of heating by the 
high energy SGR activity, although the observations were made only 2 hours 
after a soft gamma-ray burst (reported in Hurley et al. \cite{hurley99}).\\

J, H and K$'$~bands observations of all the massive stars close to the 
SGR~1806-20 position  
show no significant flux variations greater than 0.1\,magnitude 
over a time interval of four years.
Therefore, the compact source SGR~1806-20 does not form a bound binary system 
with any of these massive stars. 
This is consistent with the magnetar model (Thompson \& Duncan 
\cite{thompson}) which can explain the bursts without accretion from a 
companion donor star.\\

According to the latest results (Hurley et al. \cite{hurley99}), the SGR's 
location appears to be closer to the cluster of giant massive stars than 
to the LBV. 
Emission from the LBV's mass ejections could be superimposed on the radio 
emission of the plerion, explaining the location of this star at its radio 
core. SGR~1806-20 appears as an isolated pulsar, whose progenitor could 
have been formed as a single star or in a binary system, either with the LBV 
or in the cluster of massive stars.\\

\begin{acknowledgements}
The authors are grateful to D.A.~Frail for providing the VLA data, and to 
Richard Ogley for reading the manuscript.  
The authors are also grateful to the referee for very helpful comments.\\
\noindent The ISOCAM data presented in this paper was analyzed using ``CIA'', 
a joint development by the ESA Astrophysics Division and the ISOCAM
Consortium led by the ISOCAM PI, C. Cesarsky, Direction des Sciences de la
Mati\`ere, C.E.A., France. 
\end{acknowledgements}


\begin{thebibliography}{99}

\bibitem[1987]{atteia} Atteia,~J.-L., Boer M., Hurley K., et al., 1987, ApJ 320, L105 

\bibitem[1998]{castro-tirado} Castro-Tirado,~A.J., Gorosabel,~J., Hammersley,~P., 1998, A\&A 330, 1067 

\bibitem[1996]{cesarsky} Cesarsky,~C.J., Abergel A., Agnese P., et al., 1996, A\&A 315, L32

\bibitem[1998]{chaty} Chaty,~S., 1998, Etude multi-longueur d'onde du microquasar GRS 1915+105 et de sources binaires de haute énergie de la Galaxie, University Paris XI

\bibitem[1993]{cooke} Cooke,~B.A., 1993, Nat 366, 413

\bibitem[1997]{corbel} Corbel,~S., Wallyn P., Dame T.M., et al., 1997, ApJ 478, 624

\bibitem[1998]{dieters} Dieters,~S., Woods,~P., Kouveliotou~C., van~Paradijs,~J., 1998, IAU Circ. 6962

\bibitem[1998]{fender} Fender,~R.P., Southwell,~K., Tzioumis,~A.K., 1998, \mbox{MNRAS~298}, 692

\bibitem[1997]{frail} Frail,~D.A., Vasisht,~G., Kulkarni,~S.R., 1997, ApJ 480, L129

\bibitem[1984]{golenetskii} Golenetskii,\,S.V., Ilinskii,\,V.N., Mazets,\,E.P., 1984, Nat 307,\,41

\bibitem[1994]{hurley94} Hurley,~K., McBreen B., Rabbette M., et al., 1994, A\&A 288, L49 

\bibitem[1996]{hurley} Hurley,~K., Kouveliotou C., Fishman G.J., et al., 1996, IAU Circ. 6512

\bibitem[1999]{hurley99} Hurley,~K., Kouveliotou C., Cline T., et al., et al., 1999, ApJ 523, L37

\bibitem[1993]{kouveliotou93} Kouveliotou,~C., Morack J.M., Fishman G.J., et al., 1993, IAU Circ. 5875

\bibitem[1995]{kouveliotou95} Kouveliotou,~C., Fishman G.J., Meegan C.A., et al., 1995, IAU Circ. 6242

\bibitem[1996a]{kouveliotou96a} Kouveliotou,~C., Fishman G.J., Meegan C.A., et al., 1996a, IAU Circ. 6501

\bibitem[1996b]{kouveliotou96b} Kouveliotou,~C., van Paradijs J., Fishman G.J., et al., 1996b, IAU Circ. 6503

\bibitem[1998a]{kouveliotou98a} Kouveliotou,~C., Dieters S., Strohmayer T., et al., et al., 1998a, Nat  393, 235

\bibitem[1998b]{kouveliotou98b} Kouveliotou,~C., Kippen M., Woods P., et al., 1998b, IAU Circ. 6944

\bibitem[1999]{kouveliotou99} Kouveliotou,~C., Strohmayer T., Hurley K., et al., 1999, ApJ 510, L115

\bibitem[1993]{kulkarni93} Kulkarni,~S.R., Frail,~D.A., 1993, Nat 365, 33

\bibitem[1994]{kulkarni94} Kulkarni,~S.R., Frail,~D.A., Kassim,~N.E., Murakami,~T., Vasisht,~G., 1994, Nat 368, 129

\bibitem[1995]{kulkarni95} Kulkarni,~S.R., Matthews K., Neugebauer G., et al., 1995, ApJ 440, L61

\bibitem[1998]{kulkarni98} Kulkarni,~S.R., Thompson,~C., 1998, Nat 393, 215

\bibitem[1987]{laros} Laros,~J.G., Fenimore E.E., Klebesadel R.W., et al., 1987, ApJ 320, L111

\bibitem[1994]{murakami} Murakami,~T., Tanaka Y., Kulkarni S.R., et al., 1994, Nat 368, 127

\bibitem[1997]{ott} Ott, S., Abergel A., Altieri B., et al., 1997, Design and Implementation of CIA, the ISOCAM Interactive Analysis System. In: Hunt\,G. and Payne\,H.E. (eds.) ADASS\,VI, ASP Conference Series 125, p.34 

\bibitem[1985]{rieke} Rieke,~G.H., Lebofsky,~M.J., 1985, ApJ 288, 618

\bibitem[1997]{smith} Smith,~I.A., Schultz,~A.S.B., Hurley,~K., van~Paradijs,~J., Waters,~L.B.F.M., 1997, A\&A 319, 923

\bibitem[1993]{tanaka} Tanaka,~Y., 1993, IAU Circ. 5880

\bibitem[1995]{thompson} Thompson,~C., Duncan,~R., 1995, MNRAS 275, 255

\bibitem[1995]{van kerkwijk} van Kerkwijk,~M.H., Kulkarni,~S.R., Matthews,~K., Neugebauer,~G., 1995, ApJ 444, L33

\bibitem[1996]{van paradijs} van~Paradijs,~J., Waters,~L.B.F.M., Groot,~P.J., et al., 1996, A\&A 314, 146

\bibitem[1995]{vasisht} Vasisht,~G., Frail,~D.A., Kulkarni,~S.R., 1995, ApJ 440, L65

\end{thebibliography}
\end{document}